\documentclass[aps, prb,reprint, groupedaddress, showpacs, showkeys,twocolumn]{revtex4-1}
\usepackage{graphicx}
\usepackage{amsmath, color, ulem}
\newcommand{\dpartial}[2]{\frac{\partial #1}{\partial #2}}
\newcommand{\ve}[1]{{\bf #1}}

\newcommand{\mw}[1]{\left\langle #1 \right\rangle}

\newcommand{\mm}[0]{S^{\pm}}

\usepackage{soulutf8}
\usepackage{upgreek}
\usepackage{ulem}

\begin{document}


\title{Propagation of Thermally Induced Magnonic Spin Currents}

\author{Ulrike Ritzmann}
\author{Denise Hinzke}
\author{Ulrich Nowak}
\affiliation{Fachbereich Physik, Universit\"at Konstanz, D-78457 Konstanz, Germany}

\date{19.12.2013}

\begin{abstract}
The propagation of magnons in temperature gradients is investigated within the framework of an atomistic spin model with the stochastic Landau-Lifshitz-Gilbert equation as underlying equation of motion. We analyze the magnon accumulation, the magnon temperature profile as well as the propagation length of the excited magnons. The frequency distribution of the generated magnons is investigated in order to derive an expression for the influence of the anisotropy and the damping parameter on the magnon propagation length. For soft ferromagnetic insulators with low damping a propagation length in the range of some $\mu$m can be expected for exchange driven magnons. 
\end{abstract}

\pacs{75.30.Ds, 75.30.Sg, 75.76.+j}

\maketitle

\section{Introduction}
Spin caloritronics is a new, emerging field in magnetism describing the interplay between heat, charge and spin transport \cite{ref07,ref08}. A stimulation for this field was the discovery of the spin Seebeck effect in Permalloy by Uchida et al.\cite{ref09}. Analog to the Seebeck effect, where in an electric conductor an electrical voltage is created by applying a temperature gradient, in a ferromagnet a temperature gradient excites a spin current  leading to a spin accumulation. The generated spin accumulation was detected by measuring the spin current locally injected into a Platinum-contact using the inverse spin Hall effect\cite{ref09,ref18}. A first explanation of these effect was based on a spin-dependent Seebeck effect, where the conduction electrons propagate in two different channels and, due to a spin dependent mobility, create a spin accumulation in the system \cite{ref19}.

Interestingly, it was shown later on that this effect also appears in ferromagnetic insulators\cite{ref01}. This shows that in addition to conduction-electron spin-currents, chargeless spin-currents exist as well, where the angular momentum is transported by the magnetic excitations of the system, so-called magnons. A first theoretical description of such a magnonic spin Seebeck effect was developed by Xiao et al.\cite{ref02}. With a two temperature model including the local magnon (m) and phonon (p) temperatures the measured spin Seebeck voltage is calculated to be linearly dependent on the local difference between magnon and phonon temperature, $\Delta T_{\rm mp}=T_{\rm m}-T_{\rm p}$. This temperature difference decays with the characteristic lengthscale $\lambda$. For the ferromagnetic material YIG they estimate the length scale in the range of several millimeters.

The contribution of exchange dominated magnons to the spin Seebeck effect was investigated in recent experiments by Agrawal et al.\cite{ref11}. Using Brillouin light scattering the difference between the magnon and the phonon temperature in a system with a linear temperature gradient was determined. They found no detectable temperature difference and estimate a maximal characteristic length scale of the temperature difference of 470\,$\mu$m. One possible conclusion from this results might be be that instead of exchange magnons, magnetostatic modes mainly contribute to the spin Seebeck effect and are responsible for the long-range character of this effect.
Alternatively, phonons might contribute to the magnon accumulation as well via spin-phonon drag \cite{ref12,ref13}. 
A complete understanding of these different contributions to the spin Seebeck effect is still missing. 

In this paper thermally excited magnonic spin currents and their length scale of propagation are investigated. Using atomistic spin model simulation which describe the thermodynamics of the magnetic system in the classical limit including the whole frequency spectra of excited magnons, we describe spin currents by exchange magnons in the vicinity of a temperature step. After introducing our model, methods and basic definitions in Section II we determine the magnon accumulation as well as the corresponding magnon temperature and investigate the characteristic lengthscale of the decay of the magnon accumulation in Section III. In Section IV we introduce an analytical description which is supported by our simulations shown in Section V and gives insight into the material properties dependence of magnon propagation. 
 
\section{Magnetization profile and magnon temperature}
\label{se2}
For the investigation of magnonic spin currents in temperature gradients we use an atomistic spin model with localized spins $\ve{S}_i={\boldsymbol \mu}_i/\mu_{\rm s}$ representing the normalized magnetic moment $\mu_{\rm s}$ of a unit cell. The magnitude of the magnetic moment is assumed to be temperature independent. We simulate a three-dimensional system with simple cubic lattice structure and lattice constant $a$. The dynamics of the spin system are described in the classical limit by solving the stochastic Landau-Lifshitz-Gilbert (LLG) equation,
  \begin{align}
    \label{eq01}
    \dpartial{\ve{S}_i}{t}=-\frac{\gamma}{\mu_{\rm s}(1+\alpha^2)} \ve{S}_i\times\left(\ve{H}_i+\alpha\left(\ve{S}_i\times\ve{H}_i\right)\right)\mbox{,}
  \end{align}
numerically with the Heun method \cite{ref06} with $\gamma$ being the gyromagnetic ratio. This equation describes a precession of each spin $i$ around its effective field $\ve{H}_i$ and the coupling with the lattice by a phenomenological damping term with damping constant $\alpha$. The effective field  $\ve{H}_i$ consists of the derivative of the Hamiltonian and an additional white-noise term $\boldsymbol{\zeta}_i(t)$,
  \begin{align}
    \ve{H}_i=-\dpartial{\mathcal{H}}{\ve{S}_i}+\boldsymbol{\zeta}_i(t)\;\mbox{.}
  \end{align}
The Hamiltonian $\mathcal{H}$ in our simulation includes exchange interaction of nearest neighbors with isotropic exchange constant $J$ and an uniaxial anisotropy with an easy axis in $z$-direction and anisotropy constant $d_z$,
  \begin{align} 
    \label{eq02}
    \mathcal{H}=-\frac{J}{2} \sum_{<i,j>}{\ve{S}_i\ve{S}_j}-d_z\sum_i{S_{i,z}^2}\;\mbox{.}
  \end{align} 
The additional noise term $\boldsymbol{\zeta}_i(t)$ of the effective field $\ve{H}_i$ includes the influence of the temperature and has the following properties:
  \begin{align}
    \left\langle\boldsymbol{\zeta}(t)\right\rangle&=0\\
    \left\langle\zeta_{\eta}^i(0)\zeta_{\theta}^j(t) \right\rangle&=\frac{2k_{\rm B}T_{\rm p}\alpha\mu_{\rm s}}{\gamma}\delta_{ij}\delta_{\eta\theta}\delta (t) \;\mbox{.}
  \end{align}
Here $i,j$ denote lattice sites and $\eta$ and $\theta$ Cartesian components of the spin. 

\begin{figure}
    \includegraphics[width=0.48\textwidth]{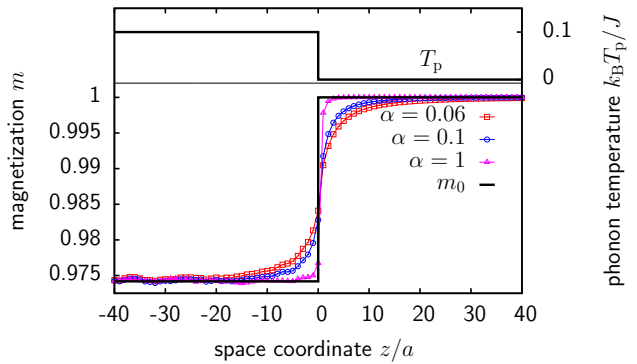}
    \caption{\label{fig1} Steady state magnetization $m$ and equilibrium magnetization $m_0$ over space coordinate $z$ for a given phonon temperature profile and for different damping parameters $\alpha$ in  a small section around the temperature step.} 
 \end{figure}

We simulate a model with a given phonon temperature $T_{\rm p}$ which is space dependent and includes a temperature step in $z$-direction in the middle of the system at $z=0$ from a temperature $T_{\rm p}^1$ in the hotter area to $T_{\rm p}^2=0\,$K (see Fig. \ref{fig1}). We assume, that this  temperature profile stays constant during the simulation and that the magnetic excitations have no influence on the phonon temperature. The system size is $8\times8\times512$, large enough to minimize finite-size effects. 

All spins are initialized parallel to the easy-axis in $z$-direction. Due to the temperature step a non-equilibrium in the magnonic density of states is created. Magnons propagate in every direction of the system, but more magnons exist in the hotter than in the colder part of the system. This leads to a constant net magnon current from the hotter towards the colder area of the system. Due to the damping of the magnons the net current appears around the temperature step with a finite length scale.After an initial relaxation time the system reaches a steady state. In this steady state the averaged spin current from the hotter towards the colder region is constant and so the local magnetization is time independent. We can now calculate the local magnetization $m(z)$  depending on the space coordinate $z$ as the time average over 
all spins in the plane perpendicular to the $z$-direction. 

We use the phonon temperature $T_{\rm p}^1=0.1J/k_{\rm B}$  in the heated area, the anisotropy constant $d_z=0.1J$ and vary the damping parameter $\alpha$. The resulting magnetization versus the space coordinate $z$ for different damping parameters in a section around the temperature step is shown in Fig. \ref{fig1}. For comparison the particular equilibrium magnetization $m_0$ of the two regions is also calculated and shown in the figure.  

 Far away from the temperature step on both sides the amplitudes of the local magnetization $m(z)$ converge to the  equilibrium values, only in the vicinity of the temperature step deviations appear. These deviations describe the magnon accumulation, induced by a surplus of magnons from the hotter region propagating towards the colder one. This leads to a less thermal excitation in the hotter area and the value of the local magnetization increases. In the colder area the surplus of incoming magnons decrease the value of the local magnetization. For smaller values of $\alpha$ the magnons can propagate over larger distances before they are finally damped. This leads to a damping-dependent magnon accumulation which increases with decreasing damping constant $\alpha$.

 \begin{figure}[b]
       \centering
    \includegraphics[width=0.48\textwidth]{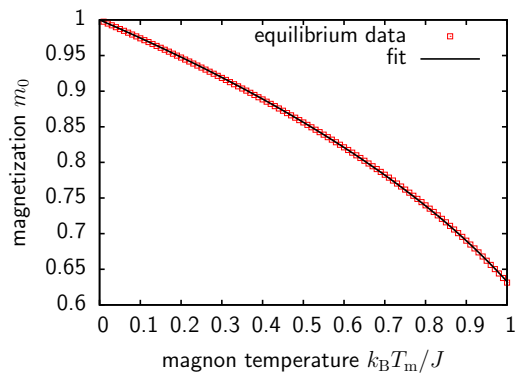}
    \caption{ \label{fig2} Equilibrium magnetization $m_0$ over the magnon temperature $T_{\rm m}$. 
    Red points show the simulated equilibrium magnetization and the black line shows a fit of the data.}
  \end{figure}
 
For a further analysis in the context of the spin-Seebeck effect we define a local magnon temperature $T_{\rm m}(z)$ via the magnetization profile $m(z)$. For that the equilibrium magnetization $m_0(T)$ is calculated for the same model but homogeneous phonon temperature $T_{\rm p}$. In equilibrium magnon temperature $T_{\rm m}$ and the phonon temperature $T_{\rm p}$ are the same and we can determine the (magnon) temperature dependence of the equilibrium magnetization $m_0(T_{\rm m})$ of the system. The magnetization of the equilibrium system decreases for increasing  magnon temperature as shown in Fig. \ref{fig2}  and the behavior can be described phenomenologically with a function \cite{ref10}
$    m_0(T) = (1 - T_{\rm m}/T_{\rm c})^{\beta}$
where $T_{\rm c}$ is the Curie temperature. Fitting our data we find $T_{\rm c}=(1.3326\pm0.00015) J/k_{\rm B}$ and for the exponent we get $\beta=0.32984\pm0.00065$. This fit of the data is also shown in Fig. \ref{fig2} and it is a good approximation over the whole temperature range. The inverse function is used in the following to determine the magnon temperature for a given local magnetization and with that a magnon temperature profile $T_{\rm m}(z)$.

\begin{figure}
    \centering
    \includegraphics[width=0.48\textwidth]{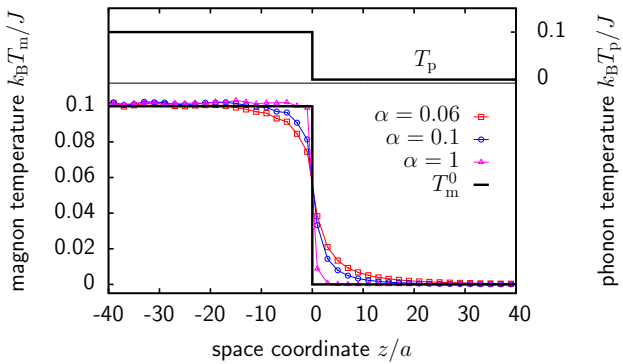}
    \caption{\label{fig3} Magnon temperature $T_{\rm m}$ over the space coordinate $z$ for different damping parameters $\alpha$ corresponding to the results in Fig. \ref{fig1}.
    }
  \end{figure}

The resulting magnon temperature profiles are shown in Fig. \ref{fig3}. Far away from the temperature step the magnon temperature $T_{\rm m}(z)$ coincides with the given phonon temperature $T_{\rm p}$, and deviations --- dependent on the damping constant $\alpha$ --- appear only around the temperature step. These deviations correspond to those of the local magnetization discussed in connection with Fig. \ref{fig1}.
    
\section{Magnon propagation length}
\label{se3}
To describe the characteristic lengthscale of the magnon propagation around the temperature step we define the magnon accumulation $\Delta m(z)$ as the difference between the relative equilibrium magnetization $m_0(z)$ at the given phonon temperature $T_{\rm p}(z)$ and the calculated local magnetization $m(z)$:
  \begin{align}
    \Delta m(z)=m_{0}(z)-m(z)\;\mbox{.}
  \end{align}
We investigate the magnon propagation in the colder part of the system, where $T_{\rm p}(z)=0$.
For a small magnon temperature, the temperature dependence of the magnetization can be approximated as
\begin{align}
   m(T_{\rm m})\approx1-\frac{\beta}{T_{\rm c}}T_{\rm m} \;\mbox{.}
 \end{align}
These linear equation is in agreement with an analytical solution for low temperatures presented by Watson et al. \cite{ref10}. For low phonon temperatures one obtains for the difference between phonon and magnon temperature
  \begin{align}
    \Delta T=T_{\rm m}-T_{\rm p}=\frac{\beta}{T_{\rm c}}\Delta m\;\mbox{.}
  \end{align}
Note, that the proportionality between magnon accumulation and temperature difference holds for higher temperatures as well as long as magnon and phonon temperature are sufficiently close so that a linear approximation applies, though the proportionality factor increases. Note also, that this proportionality was determined in theoretical descriptions of a magnonic spin Seebeck effect \cite{ref02}. Our results for the magnon accumulation should hence be relevant for the understanding of the magnonic spin Seebeck where the temperature difference between the magnons in the ferromagnet and the electrons in the non-magnet plays a key role. 

\begin{figure}
    \centering
    \includegraphics[width=0.48
    \textwidth]{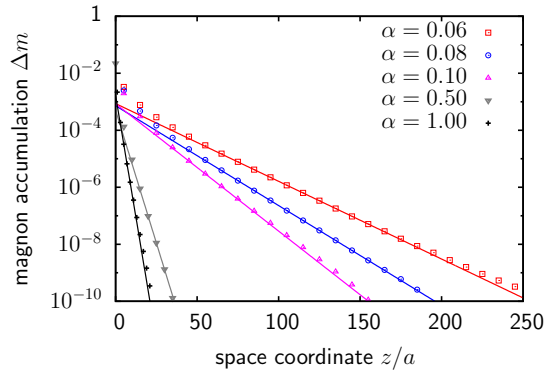}
    \caption{\label{fig4} Magnon accumulation $\Delta m$ over space coordinate $z$ in the colder region of the system at $T_{\rm p}=0$\,K for different damping constants $\alpha$ shows exponential decay with magnon propagation length $\xi$. The points show the data from our simulation and the lines the results from an exponential fit.}
  \end{figure}

We further investigate our model as before with a temperature in the heated area of $T_{\rm p}^1=0.1J/k_{\rm B}$,  anisotropy constant $d_z=0.1J$ and different damping parameters. The magnon accumulation $\Delta m$ versus the space coordinate $z$ in the colder region of the system at $T_{\rm p}=0$\,K is shown in Fig. \ref{fig4}. 
Apart from a sudden decay close to the temperature step the magnon accumulation $\Delta m(z)$ then decays exponentially on a length scale that depends on the damping constant $\alpha$. To describe this decay we fit the data with the function
  \begin{align}
    \Delta m(z)=\Delta m(0)\cdot e^{-\frac{z}{\xi}}\;\mbox{.}
  \end{align}
We define the fitting parameter $\xi$ as the propagation length of the magnons. Here, the deviations from the exponential decay at the beginning of the system are neglected. The fits for the data are shown in Fig. \ref{fig4} as continuous lines.

\begin{figure}
    \centering
    \includegraphics[width=0.48\textwidth]{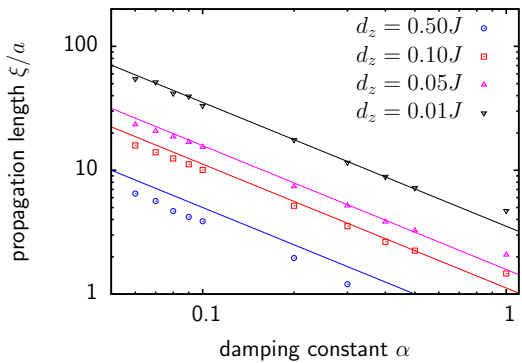}
    \caption{Magnon propagation length $\xi$ over the damping constant $\alpha$ for different anisotropy constant $d_z$. Numerical data is shown as points and the solid lines are from Eq. (\ref{eq03}).}
    \label{fig5}
  \end{figure}

The propagation length dependence on the damping parameter $\alpha$ is shown in Fig. \ref{fig5}. The values of the propagation length from our simulations, shown as points, are inversely proportional to the damping constant $\alpha$ and, furthermore, show also a strong dependence on the anisotropy constant $d_z$. This behavior will be discussed in the next two sections with an analytical analysis of the magnon propagation and an investigation of the frequencies of the propagating magnons. A simple approximation for the propagation length leads to Eq. (\ref{eq03}) which is also  shown as solid lines in Fig. \ref{fig5}.     


\section{Analytical Description with Linear Spin-Wave Theory}
\label{se4}
For the theoretical description of the magnon accumulation, excited by a temperature step in the system, we solve the LLG equation (Eq. (\ref{eq01})), analytically in the area with $T_{\rm p}=0$\,K. We consider a cubical system with lattice constant $a$ where all spins are magnetized in $z$-direction parallel to the easy-axis of the system. Assuming only small fluctuations in the $x$- and $y$-direction we have $S_i^z\approx1$ and $S_i^x,S_i^y\ll1$. 
In that case we can linearize the LLG-equation and the solution of the resulting equation consists of a sum over spin waves with wavevectors $\ve{q}$ and the related frequency $\omega_{\ve q}$ which decay exponentially in time dependent on their frequency and the damping constant $\alpha$ of the system,
  \begin{align}
   \label{eq07}
   \mm_i(t)=\frac{1}{\sqrt{N}}\sum_{\ve{q}}{\mm_{\ve{q}}(0)e^{\mp i\ve{q}\ve{r}_i\pm i\omega_{\ve{q}}t}\cdot e^{-\alpha\omega_{\ve{q}}t}}\;\mbox{.}
  \end{align}
 The frequency $\omega_{\ve q}$ of the magnons is described by the usual dispersion relation 
  \begin{align}
    \label{eq04}
    \hbar\omega_{\ve{q}}=\frac{1}{(1+\alpha^2)}\Big(2d_z+2J\sum_{\theta}(1-\cos(q_{\theta}a_{\theta})) \Big) \;\mbox{.}
  \end{align}
The dispersion relation includes a frequency gap due to the anisotropy constant and a second wavevector dependent term with a sum over the Cartesian components \cite{ref23}. 

Considering now the temperature step, magnons from the hotter area propagate towards the colder one. We investigate the damping process during that propagation in order to describe the propagating frequencies as well as to calculate the propagation length $\xi$ of the magnons for comparison with the results from section \ref{se3}. 
The magnon accumulation will depend on the distance to the temperature step and --- for small fluctuations of the $S_x$ and $S_y$ components --- can be expressed as 
  \begin{align}
    \Delta m(z)=1-\mw{S_z(z)}\approx\frac{1}{2}\mw{S_x(z)^2+S_y(z)^2}\;\mbox{,}
  \end{align}
where the brackets denote a time average.
We assume that the local fluctuations of the $S_x$ and $S_y$ components can be described with a sum over spin waves with different frequencies
and damped amplitudes $a_{\ve{q}}(z)$, 
  \begin{align}
    S_x(z)&=\sum_{\ve{q}}{a_{\ve q}(z)\cos(\omega_{\ve q}t-\ve{q}\ve{r})}\;\mbox{,}\\
    S_y(z)&=\sum_{\ve q}{a_{\ve q}(z)\sin(\omega_{\ve q}t-\ve{q}\ve{r})}\;\mbox{.}
  \end{align}
In that case for the transverse component of the magnetization one obtains
  \begin{align}
    \mw{S_x(z)^2+S_y(z)^2}=\mw{\sum_{\ve q}{a_{\ve q}(z)^2}}\;\mbox{,}
  \end{align}
where mixed terms vanish upon time averaging. The magnon accumulation can be written as:
  \begin{align}
    \label{eq12}
    \Delta m(z)=\frac{1}{2}\mw{\sum_{\ve{q}}{a_{\ve q}(z)^2}}\;\mbox{.}
  \end{align}
The amplitude $a_{\ve q}(z)$ of a magnon decays exponentially as seen in Eq. (\ref{eq07}) dependent on the damping constant and the frequency of the magnons. In the next step we describe the damping process during the propagation of the magnons. In the one-dimensional limit magnons only propagate in $z$-direction with velocity $v_q=\dpartial{\omega_{q}}{q}$. Then the propagation time can be rewritten as $t=z/v_q$ and we can describe the decay of the amplitude with $a_{\ve q}(z)=a_{\ve q}(0)\cdot f(z)$ with a damping function 
  \begin{align}
    \label{e1}
    f(z)=\exp\Big(-\frac{\alpha\omega_{\ve{q}}z}{\dpartial{\omega_{\ve{q}}}{q_z}}\Big) .
  \end{align} 

The amplitudes are damped exponentially during the propagation which defines a frequency dependent propagation length
\begin{align}
 \xi_{\omega_{\ve q}}=\frac{\sqrt{J^2- \Big(\frac{1}{2}(1+\alpha^2)( \hbar \omega_q-2d_z)-J \Big)^2}}{\alpha(1+\alpha^2)\hbar\omega_{\ve q}} ,
\end{align}
where we used $\gamma = \mu_{\rm s} / \hbar$. In the low anisotropy limit this reduces to  $\xi_{\omega_{\ve q}} = \lambda / \pi \alpha$ where $\lambda = 2 \pi / q$ is the wave length of the magnons.

The total propagation length is then the weighted average over all the excited frequencies. 
The minimal frequency is defined by the dispersion relation with a frequency gap of $\omega_q^{\rm min}=1/ (\hbar(1+\alpha^2))2d_z$. For small frequencies above that minimum the velocity is small, so the magnons are damped within  short distances. Due to the fact that the damping process is also frequency dependent higher frequencies will also be damped quickly. In the long wave length limit the minimal damping is at the frequency $\omega_q^{\rm max}\approx4d_z/(\hbar(1+\alpha^2))$ which can be determined by minimizing Eq. (\ref{e1}) .

In a three-dimensional system, besides the $z$-component of the wavevector, also transverse components of the wavevector have to be included. The damping of magnons with transverse components of the wavevector is higher than described in the one-dimensional case, because the additional transverse propagation increase the propagation time. In our simulations the cross-section is very small, so that transverse components of the wave-vectors are very high and get damped quickly. This fact and the high damping for high frequencies described in Eq. (\ref{e1}) can explain the very strong damping at the beginning of the propagation shown in Fig. \ref{fig4}.

\section{Frequencies and damping of propagating magnons}
In this section we investigate the frequency distribution of the magnonic spin current while propagating away from the temperature step. First we determine the frequencies of the propagating magnons in our simulations with Fourier transformation in time to verify our assumptions from the last section. As before a system of $8\times8\times512$ spins with a temperature step in the center of the system is simulated with an anisotropy of $d_z=0.1J$. The temperature of the heated area is $T_{\rm p}^1=0.1J/k_{\rm B}$ and the damping constant is $\alpha=0.1$. After an initial relaxation to a steady-state the frequency distribution of the propagating magnons in the colder area is determined by Fourier transformation in time of $S^{\pm}(i)=S_x(i)\pm iS_y(i)$. The frequency spectra are averaged over four points in the $x$-$y$-plane and analyzed depending on the distance $z$ of the plane to the temperature step.

\begin{figure}
   \centering
    \includegraphics[width=0.48\textwidth]{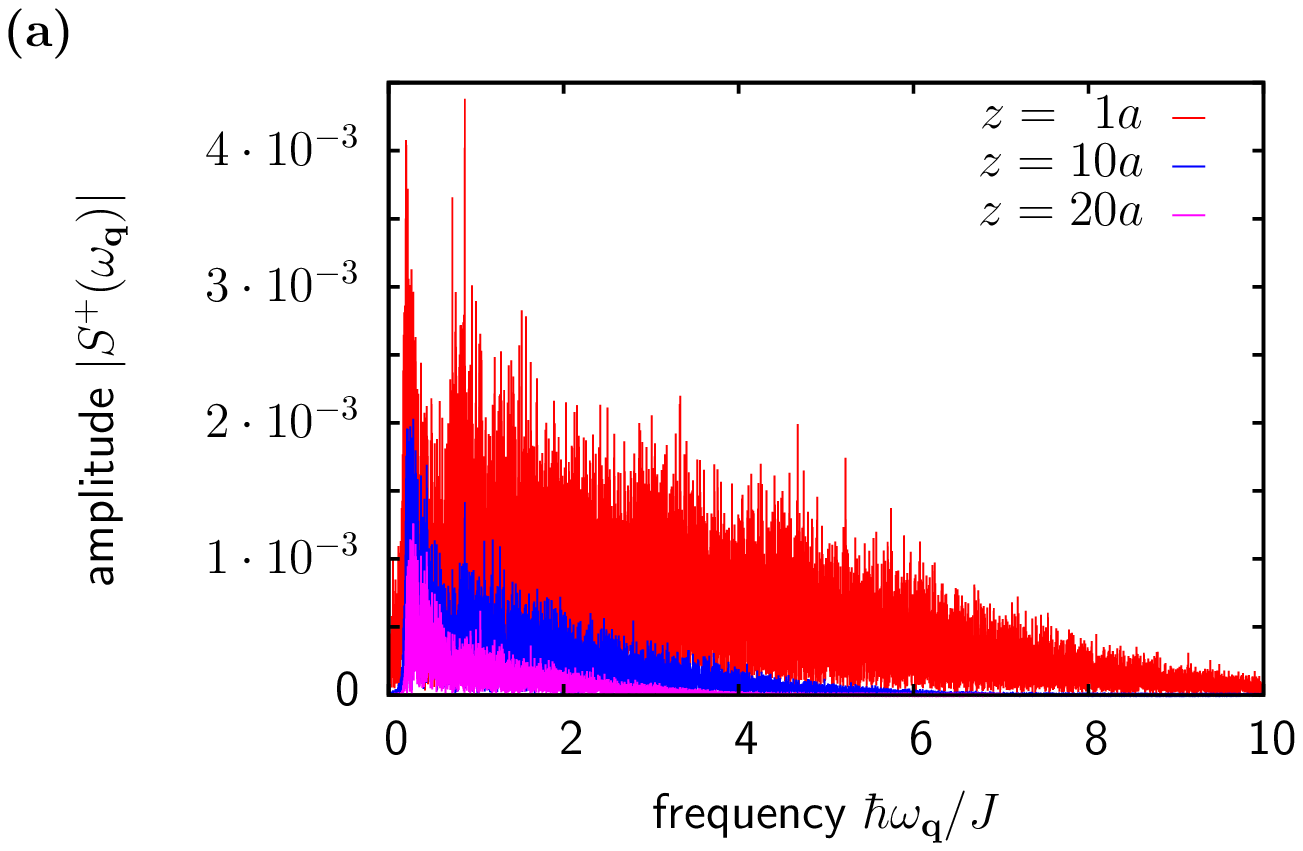}
    \includegraphics[width=0.48\textwidth]{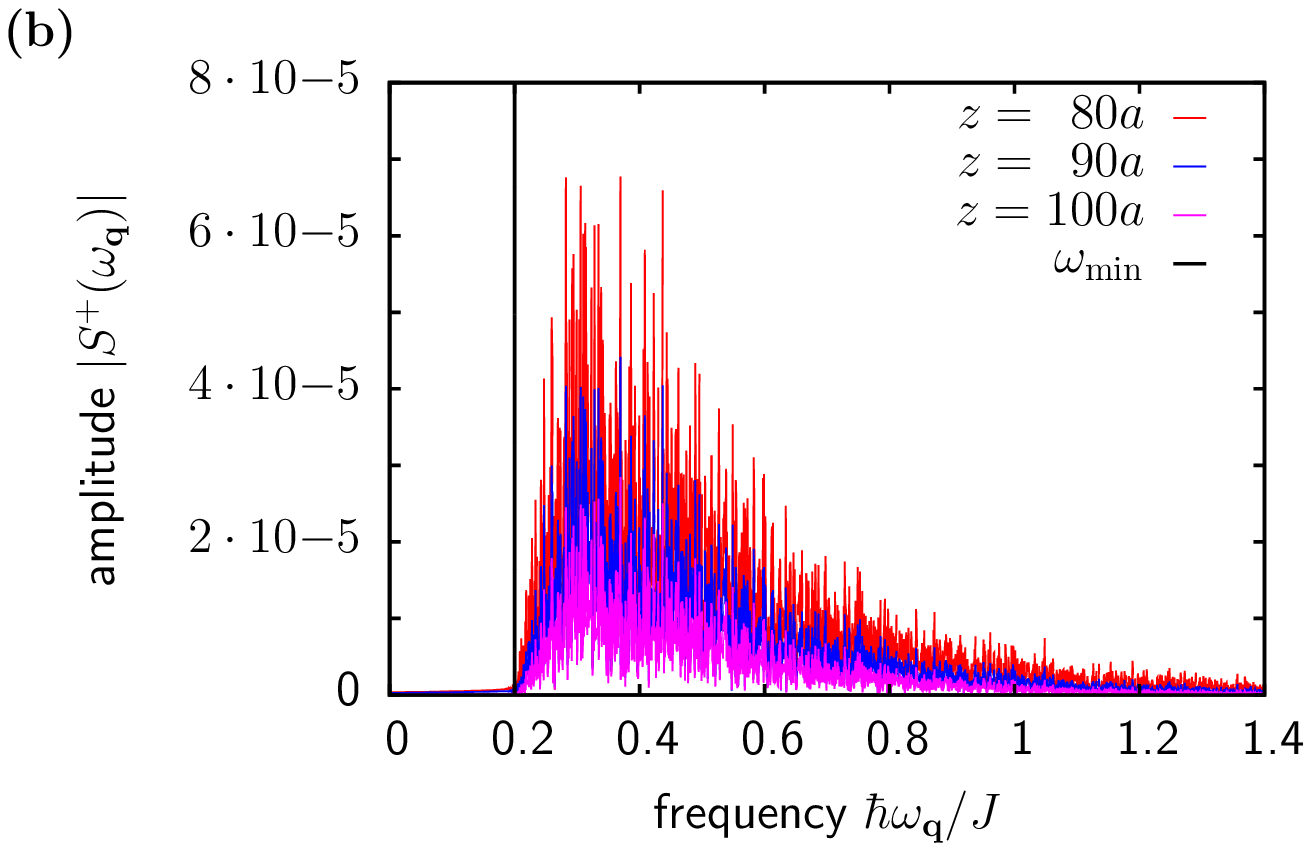}
    \caption{\label{fig9} Amplitude $|S^+(\omega_{\ve{q}})|$ versus the frequency $\omega_{\ve{q}}$ for a system with $8\times8\times512$ spins. (a):after propagation over short distances form 1 to 20 lattice constants.
    (b): after propagation over longer distances from 80 to 100 lattice constants. }
  \end{figure}

The results for small values of $z$ are shown in Fig. \ref{fig9}(a) and for higher  values of $z$, far away from the temperature step, for the regime of the exponential decay, in Fig. \ref{fig9}(b). For small values of $z$, near the temperature step, the frequency range of the propagating magnons is very broad. The minimum frequency is given by $\omega_{\ve q}^{\rm min}=2d_z/(\hbar(1+\alpha^2))$ and far away from the temperature step the maximum peak is around $\omega_{\ve q}^{\rm max}=4d_z/(\hbar(1+\alpha^2))$. These characteristic frequencies are in agreement with our findings in section \ref{se4}. 

\begin{figure}
  \centering
    \includegraphics[width=0.48\textwidth]{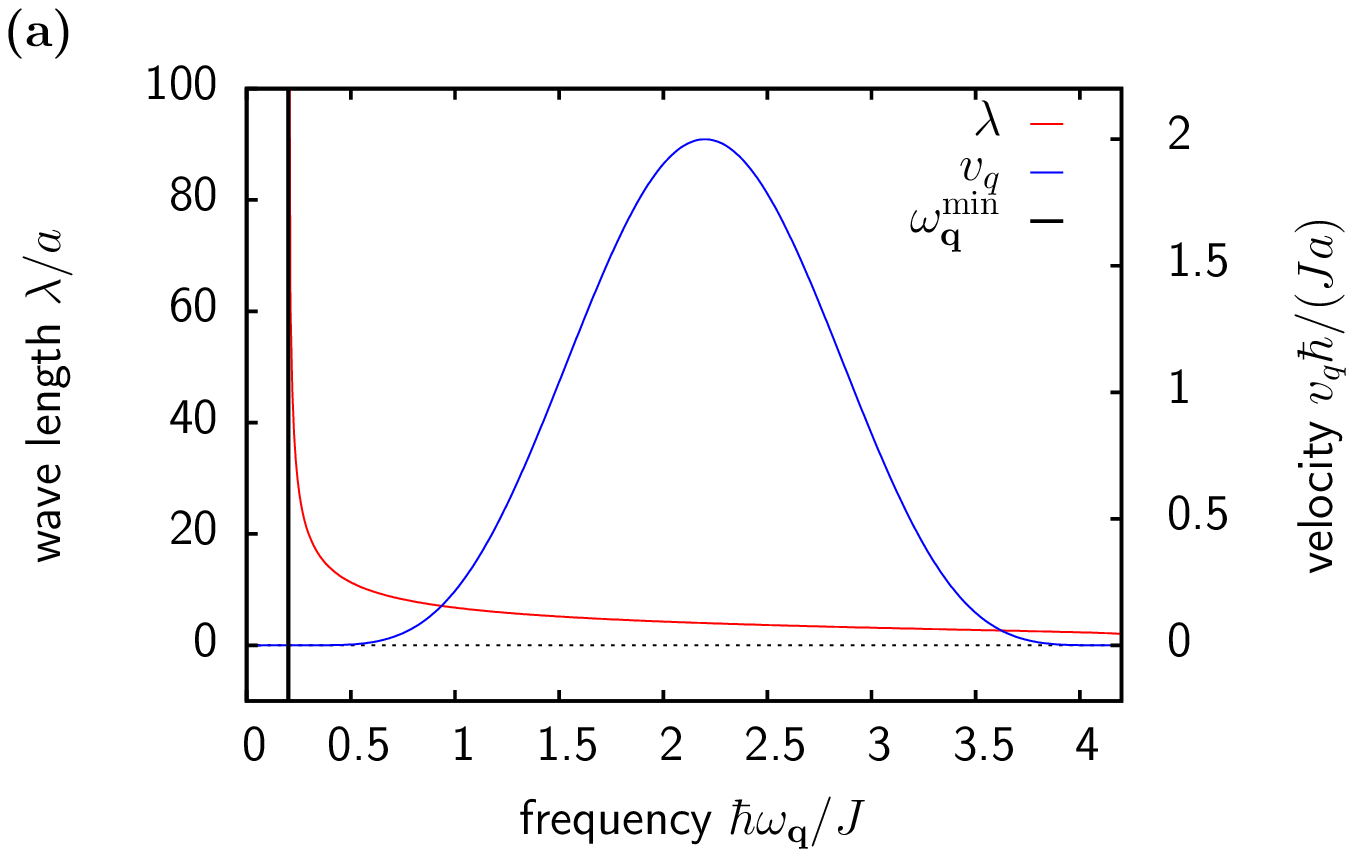}
    \includegraphics[width=0.48\textwidth]{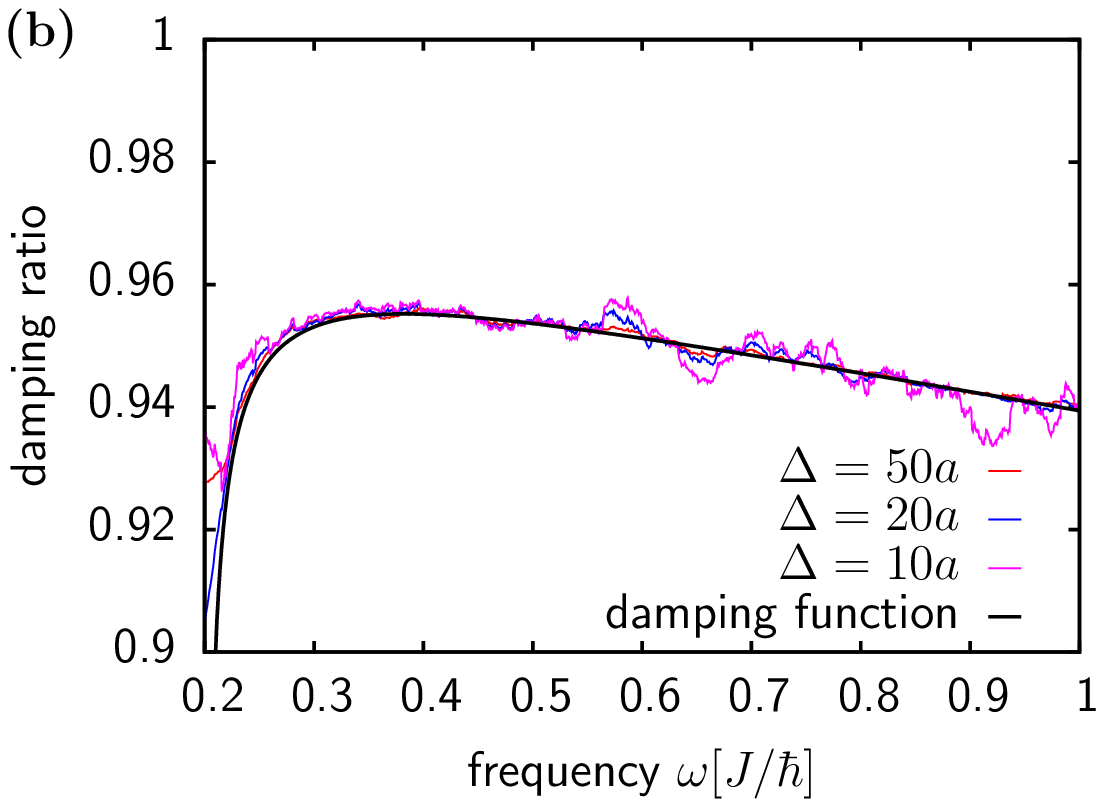}
    \caption{\label{fig10} (a): Wavelength $\lambda$ and group velocity $v_{\ve q}$ of the magnons in a one-dimensional model dependent on the frequency $\omega_{\ve q}$. (b):Damping ratio as explained in the text versus the frequency $\omega_{\ve{q}}$ for different distances $\Delta$ 
    and compared to the damping function (Eq. (\ref{e1})).}
\end{figure}

Furthermore, a stronger damping for higher frequencies can be observed. This effect corresponds to the strong damping of magnons with wavevector components transverse to the $z$-direction and it explains the higher initial damping, which was seen in the magnon accumulation in Fig. \ref{fig4}. A much narrower distribution propagates over longer distances and reaches the area shown in Fig. \ref{fig9}(b). In that area the damping can be described by one-dimensional propagation of the magnons in $z$-direction with a narrow frequency distribution around the frequency with the lowest damping  $\omega_{\ve{q}}^{\rm max}=4d_z/(\hbar(1+\alpha^2))$.The wavelength and the belonging group velocity of the magnons depending on their frequency in the one-dimensional analytical model are shown in Fig.\ref{fig10}(a). In the simulated system magnons with the longest propagation length have a wavelength of $\lambda=14a$. Depending on the ratio $d_z/J$ the wavelength increases for systems with lower anisotropy. As discussed in 
the last chapter, magnons with smaller frequencies are less damped in the time domain, but due to their smaller velocity the magnons very close to the minimum frequency also have a smaller propagation length.


To investigate the frequency-dependent damping-process during the propagation of the magnons we calculate the ratio of the amplitude of the magnons $|S^+(\omega_{\ve{q}},x)|$ for $z=80a$ and $z=80a+\Delta$ with $\Delta=10a, 20a, 50a$ and normalize it to a damping per propagation of one spin. The resulting ratios $(|S^+(\omega_{\ve{q}},x)|/|S^+(\omega_{\ve{q}},x-\Delta)|)^{1/\Delta}$ are shown in Fig. \ref{fig10} in comparison with the frequency-dependent damping-function (Eq. (\ref{e1})). The figure shows a good agreement between simulation and our analytical calculations.

These results explain the dependence of the magnon propagation length on the model parameters. The frequency with the maximal amplitude is determined by the anisotropy constant. Under the assumption that the frequency with the lowest damping is dominant and the contribution of other frequencies can be neglected the propagation length can be calculated as
  \begin{align}
    \label{eq03}
     \xi = \frac{a}{2 \alpha} \sqrt{\frac{J}{2 d_z}} \mbox{,}
  \end{align}
where the square-root term is the domain wall width of the model. This formula is also plotted in Fig. \ref{fig5}. 

The comparison with our simulations shows good agreement though the  equation above gives only the propagation of those magnons with the smallest damping during the propagation.In the considered system with $\alpha=0.1$ and $d_z=0.1J$ we get a propagation length of about $\xi=11a$ at a wavelength of the magnons $\lambda=14a$. For smaller values of the anisotropy and smaller damping parameters the frequency distribution of the thermal magnons is broader and Eq. (\ref{eq03}) is an overestimation of the real propagation length since the magnon accumulation is no longer exponentially decaying due to the broader spectrum of propagating frequencies. However we would expect for soft ferromagnetic insulators with a small damping constant of $10^{-4}-10^{-3}$ and an anisotropy constant in the range of $10^{-3}J-10^{-2}J$ a propagation length of $10^3a-10^5a$ which would be in the micrometer-range.

\section{Summary and Discussion}
Using the framework of an atomistic spin model we describe thermally induced magnon propagation in a model containing  a temperature step. The results give an impression of the relevant length scale of the propagation of thermally induced exchange magnons and its dependence on system parameters as the anisotropy, the exchange and the damping constant. In the heated area magnons with a broad frequency distribution are generated and because of the very strong damping for magnons with high frequency, especially those with wave-vector components transverse to the propagation direction in $z$-direction, most of the induced magnons are damped on shorter length scales. Behind this region of strong damping near the temperature step, the propagation of  magnons is unidirectional and the magnon accumulation decays exponentially with the characteristic propagation length $\xi$. This propagation length depends on the damping parameter but also on system properties as the anisotropy of the system, because of the 
dependence on the induced frequencies.

In contrast to long range magnetostatic spin waves, which can propagate over distances of some mm \cite{ref16, ref25}, we find that for exchange magnons the propagation length is considerably shorter and expect from our findings for soft ferromagnetic insulators with a low damping constant a propagation length in the range of some $\mu$m for those magnons close to the frequency gap and the lowest damping. These findings will contribute to the understanding of length scale dependent investigations of the spin Seebeck effect \cite{ref11,ref24,ref30,ref31}.

Recent experiments investigate the longitudinal spin Seebeck effect, where the generated spin current longitudinal to the applied temperature gradient is measured \cite{ref28,ref29,ref26,ref27}. In this configuration Kehlberger et al. show that the measured spin current is dependent on the thickness of the YIG layer and they observe a saturation of the spin current on a lengthscale of 100\;nm \cite{ref24}. This saturation can be explained by the lengthscale of the propagation of the thermally excited magnons. Only those magnons reaching the YIG/Pt interface of the sample contribute to the measured spin current and --- as shown here --- exchange magnons thermally excited at larger distances are damped before they can reach the interface. In this paper, we focus on the propagation length of those magnons with the lowest damping, however the lengthscale of the magnon accumulation 
at the end of a temperature gradient is dominated by a broad range of magnons with higher frequencies which are therefore damped on 
shorter length scales.


%


\begin{acknowledgments}
 The authors would like to thank the Deutsche Forschungsgemeinschaft (DFG) for financial support via SPP 1538 ``Spin Caloric Transport'' and the SFB 767 ``Controlled Nanosystem: Interaction and Interfacing to the Macroscale''.
\end{acknowledgments}

\bibliography{paper.bib}

\begin{thebibliography}{22}%
\makeatletter
\providecommand \@ifxundefined [1]{%
 \@ifx{#1\undefined}
}%
\providecommand \@ifnum [1]{%
 \ifnum #1\expandafter \@firstoftwo
 \else \expandafter \@secondoftwo
 \fi
}%
\providecommand \@ifx [1]{%
 \ifx #1\expandafter \@firstoftwo
 \else \expandafter \@secondoftwo
 \fi
}%
\providecommand \natexlab [1]{#1}%
\providecommand \enquote  [1]{``#1''}%
\providecommand \bibnamefont  [1]{#1}%
\providecommand \bibfnamefont [1]{#1}%
\providecommand \citenamefont [1]{#1}%
\providecommand \href@noop [0]{\@secondoftwo}%
\providecommand \href [0]{\begingroup \@sanitize@url \@href}%
\providecommand \@href[1]{\@@startlink{#1}\@@href}%
\providecommand \@@href[1]{\endgroup#1\@@endlink}%
\providecommand \@sanitize@url [0]{\catcode `\\12\catcode `\$12\catcode
  `\&12\catcode `\#12\catcode `\^12\catcode `\_12\catcode `\%12\relax}%
\providecommand \@@startlink[1]{}%
\providecommand \@@endlink[0]{}%
\providecommand \url  [0]{\begingroup\@sanitize@url \@url }%
\providecommand \@url [1]{\endgroup\@href {#1}{\urlprefix }}%
\providecommand \urlprefix  [0]{URL }%
\providecommand \Eprint [0]{\href }%
\providecommand \doibase [0]{http://dx.doi.org/}%
\providecommand \selectlanguage [0]{\@gobble}%
\providecommand \bibinfo  [0]{\@secondoftwo}%
\providecommand \bibfield  [0]{\@secondoftwo}%
\providecommand \translation [1]{[#1]}%
\providecommand \BibitemOpen [0]{}%
\providecommand \bibitemStop [0]{}%
\providecommand \bibitemNoStop [0]{.\EOS\space}%
\providecommand \EOS [0]{\spacefactor3000\relax}%
\providecommand \BibitemShut  [1]{\csname bibitem#1\endcsname}%
\let\auto@bib@innerbib\@empty
\bibitem [{\citenamefont {Bauer}\ \emph {et~al.}(2012)\citenamefont {Bauer},
  \citenamefont {Saitoh},\ and\ \citenamefont {van Wees}}]{ref07}%
  \BibitemOpen
  \bibfield  {author} {\bibinfo {author} {\bibfnamefont {G.~E.~W.}\
  \bibnamefont {Bauer}}, \bibinfo {author} {\bibfnamefont {E.}~\bibnamefont
  {Saitoh}}, \ and\ \bibinfo {author} {\bibfnamefont {B.~J.}\ \bibnamefont {van
  Wees}},\ }\href {http://dx.doi.org/10.1038/nmat3301} {\bibfield  {journal}
  {\bibinfo  {journal} {Nature Mater.}\ }\textbf {\bibinfo {volume} {11}},\
  \bibinfo {pages} {391} (\bibinfo {year} {2012})}\BibitemShut {NoStop}%
\bibitem [{\citenamefont {Bauer}\ \emph {et~al.}(2010)\citenamefont {Bauer},
  \citenamefont {MacDonald},\ and\ \citenamefont {Maekawa}}]{ref08}%
  \BibitemOpen
  \bibfield  {author} {\bibinfo {author} {\bibfnamefont {G.~E.~W.}\
  \bibnamefont {Bauer}}, \bibinfo {author} {\bibfnamefont {A.~H.}\ \bibnamefont
  {MacDonald}}, \ and\ \bibinfo {author} {\bibfnamefont {S.}~\bibnamefont
  {Maekawa}},\ }\href {\doibase DOI: 10.1016/j.ssc.2010.01.022} {\bibfield
  {journal} {\bibinfo  {journal} {Solid State Commun.}\ }\textbf {\bibinfo
  {volume} {150}},\ \bibinfo {pages} {459 } (\bibinfo {year}
  {2010})}\BibitemShut {NoStop}%
\bibitem [{\citenamefont {Uchida}\ \emph {et~al.}(2008)\citenamefont {Uchida},
  \citenamefont {Takahashi}, \citenamefont {Harii}, \citenamefont {Ieda},
  \citenamefont {Koshibae}, \citenamefont {Ando}, \citenamefont {Maekawa},\
  and\ \citenamefont {Saitoh}}]{ref09}%
  \BibitemOpen
  \bibfield  {author} {\bibinfo {author} {\bibfnamefont {K.}~\bibnamefont
  {Uchida}}, \bibinfo {author} {\bibfnamefont {S.}~\bibnamefont {Takahashi}},
  \bibinfo {author} {\bibfnamefont {K.}~\bibnamefont {Harii}}, \bibinfo
  {author} {\bibfnamefont {J.}~\bibnamefont {Ieda}}, \bibinfo {author}
  {\bibfnamefont {W.}~\bibnamefont {Koshibae}}, \bibinfo {author}
  {\bibfnamefont {K.}~\bibnamefont {Ando}}, \bibinfo {author} {\bibfnamefont
  {S.}~\bibnamefont {Maekawa}}, \ and\ \bibinfo {author} {\bibfnamefont
  {E.}~\bibnamefont {Saitoh}},\ }\href {\doibase 10.1038/nature07321}
  {\bibfield  {journal} {\bibinfo  {journal} {Nature}\ }\textbf {\bibinfo
  {volume} {455}},\ \bibinfo {pages} {778} (\bibinfo {year}
  {2008})}\BibitemShut {NoStop}%
\bibitem [{\citenamefont {Saitoh}\ \emph {et~al.}(2006)\citenamefont {Saitoh},
  \citenamefont {Ueda}, \citenamefont {Miyajima},\ and\ \citenamefont
  {Tatara}}]{ref18}%
  \BibitemOpen
  \bibfield  {author} {\bibinfo {author} {\bibfnamefont {E.}~\bibnamefont
  {Saitoh}}, \bibinfo {author} {\bibfnamefont {M.}~\bibnamefont {Ueda}},
  \bibinfo {author} {\bibfnamefont {H.}~\bibnamefont {Miyajima}}, \ and\
  \bibinfo {author} {\bibfnamefont {G.}~\bibnamefont {Tatara}},\ }\href
  {\doibase 10.1063/1.2199473} {\bibfield  {journal} {\bibinfo  {journal}
  {Appl. Phys. Lett.}\ }\textbf {\bibinfo {volume} {88}},\ \bibinfo {eid}
  {182509} (\bibinfo {year} {2006})}\BibitemShut {NoStop}%
\bibitem [{\citenamefont {Uchida}\ \emph {et~al.}(2009)\citenamefont {Uchida},
  \citenamefont {Takahashi}, \citenamefont {Ieda}, \citenamefont {Harii},
  \citenamefont {Ikeda}, \citenamefont {Koshibae}, \citenamefont {Maekawa},\
  and\ \citenamefont {Saitoh}}]{ref19}%
  \BibitemOpen
  \bibfield  {author} {\bibinfo {author} {\bibfnamefont {K.}~\bibnamefont
  {Uchida}}, \bibinfo {author} {\bibfnamefont {S.}~\bibnamefont {Takahashi}},
  \bibinfo {author} {\bibfnamefont {J.}~\bibnamefont {Ieda}}, \bibinfo {author}
  {\bibfnamefont {K.}~\bibnamefont {Harii}}, \bibinfo {author} {\bibfnamefont
  {K.}~\bibnamefont {Ikeda}}, \bibinfo {author} {\bibfnamefont
  {W.}~\bibnamefont {Koshibae}}, \bibinfo {author} {\bibfnamefont
  {S.}~\bibnamefont {Maekawa}}, \ and\ \bibinfo {author} {\bibfnamefont
  {E.}~\bibnamefont {Saitoh}},\ }\href {\doibase 10.1063/1.3056581} {\bibfield
  {journal} {\bibinfo  {journal} {J. Appl. Phys.}\ }\textbf {\bibinfo {volume}
  {105}},\ \bibinfo {eid} {07C908} (\bibinfo {year} {2009})}\BibitemShut
  {NoStop}%
\bibitem [{\citenamefont {Uchida}\ \emph
  {et~al.}(2010{\natexlab{a}})\citenamefont {Uchida}, \citenamefont {Xiao},
  \citenamefont {Adachi}, \citenamefont {Ohe}, \citenamefont {Takahashi},
  \citenamefont {Ieda}, \citenamefont {Ota}, \citenamefont {Kajiwara},
  \citenamefont {Umezawa}, \citenamefont {Kawai}, \citenamefont {Bauer},
  \citenamefont {Maekawa},\ and\ \citenamefont {Saitoh}}]{ref01}%
  \BibitemOpen
  \bibfield  {author} {\bibinfo {author} {\bibfnamefont {K.}~\bibnamefont
  {Uchida}}, \bibinfo {author} {\bibfnamefont {J.}~\bibnamefont {Xiao}},
  \bibinfo {author} {\bibfnamefont {H.}~\bibnamefont {Adachi}}, \bibinfo
  {author} {\bibfnamefont {J.}~\bibnamefont {Ohe}}, \bibinfo {author}
  {\bibfnamefont {S.}~\bibnamefont {Takahashi}}, \bibinfo {author}
  {\bibfnamefont {J.}~\bibnamefont {Ieda}}, \bibinfo {author} {\bibfnamefont
  {T.}~\bibnamefont {Ota}}, \bibinfo {author} {\bibfnamefont {Y.}~\bibnamefont
  {Kajiwara}}, \bibinfo {author} {\bibfnamefont {H.}~\bibnamefont {Umezawa}},
  \bibinfo {author} {\bibfnamefont {H.}~\bibnamefont {Kawai}}, \bibinfo
  {author} {\bibfnamefont {G.~E.~W.}\ \bibnamefont {Bauer}}, \bibinfo {author}
  {\bibfnamefont {S.}~\bibnamefont {Maekawa}}, \ and\ \bibinfo {author}
  {\bibfnamefont {E.}~\bibnamefont {Saitoh}},\ }\href {\doibase
  10.1038/nmat2856} {\bibfield  {journal} {\bibinfo  {journal} {Nature Mater.}\
  }\textbf {\bibinfo {volume} {9}},\ \bibinfo {pages} {894} (\bibinfo {year}
  {2010}{\natexlab{a}})}\BibitemShut {NoStop}%
\bibitem [{\citenamefont {Xiao}\ \emph {et~al.}(2010)\citenamefont {Xiao},
  \citenamefont {Bauer}, \citenamefont {Uchida}, \citenamefont {Saitoh},\ and\
  \citenamefont {Maekawa}}]{ref02}%
  \BibitemOpen
  \bibfield  {author} {\bibinfo {author} {\bibfnamefont {J.}~\bibnamefont
  {Xiao}}, \bibinfo {author} {\bibfnamefont {G.~E.~W.}\ \bibnamefont {Bauer}},
  \bibinfo {author} {\bibfnamefont {K.-c.}\ \bibnamefont {Uchida}}, \bibinfo
  {author} {\bibfnamefont {E.}~\bibnamefont {Saitoh}}, \ and\ \bibinfo {author}
  {\bibfnamefont {S.}~\bibnamefont {Maekawa}},\ }\href {\doibase
  10.1103/PhysRevB.81.214418} {\bibfield  {journal} {\bibinfo  {journal} {Phys.
  Rev. B}\ }\textbf {\bibinfo {volume} {81}},\ \bibinfo {pages} {214418}
  (\bibinfo {year} {2010})}\BibitemShut {NoStop}%
\bibitem [{\citenamefont {Agrawal}\ \emph {et~al.}(2013)\citenamefont
  {Agrawal}, \citenamefont {Vasyuchka}, \citenamefont {Serga}, \citenamefont
  {Karenowska}, \citenamefont {Melkov},\ and\ \citenamefont
  {Hillebrands}}]{ref11}%
  \BibitemOpen
  \bibfield  {author} {\bibinfo {author} {\bibfnamefont {M.}~\bibnamefont
  {Agrawal}}, \bibinfo {author} {\bibfnamefont {V.~I.}\ \bibnamefont
  {Vasyuchka}}, \bibinfo {author} {\bibfnamefont {A.~A.}\ \bibnamefont
  {Serga}}, \bibinfo {author} {\bibfnamefont {A.~D.}\ \bibnamefont
  {Karenowska}}, \bibinfo {author} {\bibfnamefont {G.~A.}\ \bibnamefont
  {Melkov}}, \ and\ \bibinfo {author} {\bibfnamefont {B.}~\bibnamefont
  {Hillebrands}},\ }\href {\doibase 10.1103/PhysRevLett.111.107204} {\bibfield
  {journal} {\bibinfo  {journal} {Phys. Rev. Lett.}\ }\textbf {\bibinfo
  {volume} {111}},\ \bibinfo {pages} {107204} (\bibinfo {year}
  {2013})}\BibitemShut {NoStop}%
\bibitem [{\citenamefont {Adachi}\ \emph {et~al.}(2010)\citenamefont {Adachi},
  \citenamefont {ichi Uchida}, \citenamefont {Saitoh}, \citenamefont {ichiro
  Ohe}, \citenamefont {Takahashi},\ and\ \citenamefont {Maekawa}}]{ref12}%
  \BibitemOpen
  \bibfield  {author} {\bibinfo {author} {\bibfnamefont {H.}~\bibnamefont
  {Adachi}}, \bibinfo {author} {\bibfnamefont {K.}~\bibnamefont {ichi Uchida}},
  \bibinfo {author} {\bibfnamefont {E.}~\bibnamefont {Saitoh}}, \bibinfo
  {author} {\bibfnamefont {J.}~\bibnamefont {ichiro Ohe}}, \bibinfo {author}
  {\bibfnamefont {S.}~\bibnamefont {Takahashi}}, \ and\ \bibinfo {author}
  {\bibfnamefont {S.}~\bibnamefont {Maekawa}},\ }\href {\doibase
  10.1063/1.3529944} {\bibfield  {journal} {\bibinfo  {journal} {Appl.Phys.
  Lett.}\ }\textbf {\bibinfo {volume} {97}},\ \bibinfo {eid} {252506} (\bibinfo
  {year} {2010})}\BibitemShut {NoStop}%
\bibitem [{\citenamefont {Adachi}\ \emph {et~al.}(2011)\citenamefont {Adachi},
  \citenamefont {Ohe}, \citenamefont {Takahashi},\ and\ \citenamefont
  {Maekawa}}]{ref13}%
  \BibitemOpen
  \bibfield  {author} {\bibinfo {author} {\bibfnamefont {H.}~\bibnamefont
  {Adachi}}, \bibinfo {author} {\bibfnamefont {J.-i.}\ \bibnamefont {Ohe}},
  \bibinfo {author} {\bibfnamefont {S.}~\bibnamefont {Takahashi}}, \ and\
  \bibinfo {author} {\bibfnamefont {S.}~\bibnamefont {Maekawa}},\ }\href
  {\doibase 10.1103/PhysRevB.83.094410} {\bibfield  {journal} {\bibinfo
  {journal} {Phys. Rev. B}\ }\textbf {\bibinfo {volume} {83}},\ \bibinfo
  {pages} {094410} (\bibinfo {year} {2011})}\BibitemShut {NoStop}%
\bibitem [{\citenamefont {Nowak}(2007)}]{ref06}%
  \BibitemOpen
  \bibfield  {author} {\bibinfo {author} {\bibfnamefont {U.}~\bibnamefont
  {Nowak}},\ }\enquote {\bibinfo {title} {Handbook of magnetism and advanced
  magnetic materials},}\ \ (\bibinfo  {publisher} {John Wiley \& Sons},\
  \bibinfo {year} {2007})\ Chap.\ \bibinfo {chapter} {Classical
  Spin-Models}\BibitemShut {NoStop}%
\bibitem [{\citenamefont {Watson}\ \emph {et~al.}(1969)\citenamefont {Watson},
  \citenamefont {Blume},\ and\ \citenamefont {Vineyard}}]{ref10}%
  \BibitemOpen
  \bibfield  {author} {\bibinfo {author} {\bibfnamefont {R.~E.}\ \bibnamefont
  {Watson}}, \bibinfo {author} {\bibfnamefont {M.}~\bibnamefont {Blume}}, \
  and\ \bibinfo {author} {\bibfnamefont {G.~H.}\ \bibnamefont {Vineyard}},\
  }\href {\doibase 10.1103/PhysRev.181.811} {\bibfield  {journal} {\bibinfo
  {journal} {Phys. Rev.}\ }\textbf {\bibinfo {volume} {181}},\ \bibinfo {pages}
  {811} (\bibinfo {year} {1969})}\BibitemShut {NoStop}%
\bibitem [{\citenamefont {Atxitia}\ \emph {et~al.}(2010)\citenamefont
  {Atxitia}, \citenamefont {Hinzke}, \citenamefont {Chubykalo-Fesenko},
  \citenamefont {Nowak}, \citenamefont {Kachkachi}, \citenamefont {Mryasov},
  \citenamefont {Evans},\ and\ \citenamefont {Chantrell}}]{ref23}%
  \BibitemOpen
  \bibfield  {author} {\bibinfo {author} {\bibfnamefont {U.}~\bibnamefont
  {Atxitia}}, \bibinfo {author} {\bibfnamefont {D.}~\bibnamefont {Hinzke}},
  \bibinfo {author} {\bibfnamefont {O.}~\bibnamefont {Chubykalo-Fesenko}},
  \bibinfo {author} {\bibfnamefont {U.}~\bibnamefont {Nowak}}, \bibinfo
  {author} {\bibfnamefont {H.}~\bibnamefont {Kachkachi}}, \bibinfo {author}
  {\bibfnamefont {O.~N.}\ \bibnamefont {Mryasov}}, \bibinfo {author}
  {\bibfnamefont {R.~F.}\ \bibnamefont {Evans}}, \ and\ \bibinfo {author}
  {\bibfnamefont {R.~W.}\ \bibnamefont {Chantrell}},\ }\href {\doibase
  10.1103/PhysRevB.82.134440} {\bibfield  {journal} {\bibinfo  {journal} {Phys.
  Rev. B}\ }\textbf {\bibinfo {volume} {82}},\ \bibinfo {pages} {134440}
  (\bibinfo {year} {2010})}\BibitemShut {NoStop}%
\bibitem [{\citenamefont {Kajiwara}\ \emph {et~al.}(2010)\citenamefont
  {Kajiwara}, \citenamefont {Harii}, \citenamefont {Takahashi}, \citenamefont
  {Ohe}, \citenamefont {Uchida}, \citenamefont {Mizuguchi}, \citenamefont
  {Umezawa}, \citenamefont {Kawai}, \citenamefont {Ando}, \citenamefont
  {Takanashi}, \citenamefont {Maekawa},\ and\ \citenamefont {Saitoh}}]{ref16}%
  \BibitemOpen
  \bibfield  {author} {\bibinfo {author} {\bibfnamefont {Y.}~\bibnamefont
  {Kajiwara}}, \bibinfo {author} {\bibfnamefont {K.}~\bibnamefont {Harii}},
  \bibinfo {author} {\bibfnamefont {S.}~\bibnamefont {Takahashi}}, \bibinfo
  {author} {\bibfnamefont {J.}~\bibnamefont {Ohe}}, \bibinfo {author}
  {\bibfnamefont {K.}~\bibnamefont {Uchida}}, \bibinfo {author} {\bibfnamefont
  {M.}~\bibnamefont {Mizuguchi}}, \bibinfo {author} {\bibfnamefont
  {H.}~\bibnamefont {Umezawa}}, \bibinfo {author} {\bibfnamefont
  {H.}~\bibnamefont {Kawai}}, \bibinfo {author} {\bibfnamefont
  {K.}~\bibnamefont {Ando}}, \bibinfo {author} {\bibfnamefont {K.}~\bibnamefont
  {Takanashi}}, \bibinfo {author} {\bibfnamefont {S.}~\bibnamefont {Maekawa}},
  \ and\ \bibinfo {author} {\bibfnamefont {E.}~\bibnamefont {Saitoh}},\ }\href
  {\doibase 10.1038/nature08876} {\bibfield  {journal} {\bibinfo  {journal}
  {Nature}\ }\textbf {\bibinfo {volume} {464}},\ \bibinfo {pages} {262}
  (\bibinfo {year} {2010})}\BibitemShut {NoStop}%
\bibitem [{\citenamefont {An}\ \emph {et~al.}(2013)\citenamefont {An},
  \citenamefont {Vasyuchka}, \citenamefont {Uchida}, \citenamefont {Chumak},
  \citenamefont {Yamaguchi}, \citenamefont {Harii}, \citenamefont {Ohe},
  \citenamefont {Jungfleisch}, \citenamefont {Kajiwara}, \citenamefont
  {Adachi}, \citenamefont {Hillebrands}, \citenamefont {Maekawa},\ and\
  \citenamefont {Saitoh}}]{ref25}%
  \BibitemOpen
  \bibfield  {author} {\bibinfo {author} {\bibfnamefont {T.}~\bibnamefont
  {An}}, \bibinfo {author} {\bibfnamefont {V.~I.}\ \bibnamefont {Vasyuchka}},
  \bibinfo {author} {\bibfnamefont {K.}~\bibnamefont {Uchida}}, \bibinfo
  {author} {\bibfnamefont {A.~V.}\ \bibnamefont {Chumak}}, \bibinfo {author}
  {\bibfnamefont {K.}~\bibnamefont {Yamaguchi}}, \bibinfo {author}
  {\bibfnamefont {K.}~\bibnamefont {Harii}}, \bibinfo {author} {\bibfnamefont
  {J.}~\bibnamefont {Ohe}}, \bibinfo {author} {\bibfnamefont {M.~B.}\
  \bibnamefont {Jungfleisch}}, \bibinfo {author} {\bibfnamefont
  {Y.}~\bibnamefont {Kajiwara}}, \bibinfo {author} {\bibfnamefont
  {H.}~\bibnamefont {Adachi}}, \bibinfo {author} {\bibfnamefont
  {B.}~\bibnamefont {Hillebrands}}, \bibinfo {author} {\bibfnamefont
  {S.}~\bibnamefont {Maekawa}}, \ and\ \bibinfo {author} {\bibfnamefont
  {E.}~\bibnamefont {Saitoh}},\ }\href {http://dx.doi.org/10.1038/nmat3628}
  {\bibfield  {journal} {\bibinfo  {journal} {Nature Mater.}\ }\textbf
  {\bibinfo {volume} {12}},\ \bibinfo {pages} {549} (\bibinfo {year}
  {2013})}\BibitemShut {NoStop}%
\bibitem [{\citenamefont {Kehlberger}\ \emph {et~al.}()\citenamefont
  {Kehlberger}, \citenamefont {R\"oser}, \citenamefont {Jacob}, \citenamefont
  {Ritzmann}, \citenamefont {Hinzke}, \citenamefont {Nowak}, \citenamefont
  {Ombasli}, \citenamefont {Kim}, \citenamefont {Ross}, \citenamefont
  {Jungfleisch}, \citenamefont {Hillebrands},\ and\ \citenamefont
  {Kl\"aui}}]{ref24}%
  \BibitemOpen
  \bibfield  {author} {\bibinfo {author} {\bibfnamefont {A.}~\bibnamefont
  {Kehlberger}}, \bibinfo {author} {\bibfnamefont {R.}~\bibnamefont {R\"oser}},
  \bibinfo {author} {\bibfnamefont {G.}~\bibnamefont {Jacob}}, \bibinfo
  {author} {\bibfnamefont {U.}~\bibnamefont {Ritzmann}}, \bibinfo {author}
  {\bibfnamefont {D.}~\bibnamefont {Hinzke}}, \bibinfo {author} {\bibfnamefont
  {U.}~\bibnamefont {Nowak}}, \bibinfo {author} {\bibfnamefont
  {M.}~\bibnamefont {Ombasli}}, \bibinfo {author} {\bibfnamefont {D.~H.}\
  \bibnamefont {Kim}}, \bibinfo {author} {\bibfnamefont {C.~A.}\ \bibnamefont
  {Ross}}, \bibinfo {author} {\bibfnamefont {M.~B.}\ \bibnamefont
  {Jungfleisch}}, \bibinfo {author} {\bibfnamefont {B.}~\bibnamefont
  {Hillebrands}}, \ and\ \bibinfo {author} {\bibfnamefont {M.}~\bibnamefont
  {Kl\"aui}},\ }\href@noop {} {\bibinfo  {journal} {arXiv: 1306.0784}\
  }\BibitemShut {NoStop}%
\bibitem [{\citenamefont {Hoffman}\ \emph {et~al.}(2013)\citenamefont
  {Hoffman}, \citenamefont {Sato},\ and\ \citenamefont {Tserkovnyak}}]{ref30}%
  \BibitemOpen
\bibfield  {journal} {  }\bibfield  {author} {\bibinfo {author} {\bibfnamefont
  {S.}~\bibnamefont {Hoffman}}, \bibinfo {author} {\bibfnamefont
  {K.}~\bibnamefont {Sato}}, \ and\ \bibinfo {author} {\bibfnamefont
  {Y.}~\bibnamefont {Tserkovnyak}},\ }\href {\doibase
  10.1103/PhysRevB.88.064408} {\bibfield  {journal} {\bibinfo  {journal} {Phys.
  Rev. B}\ }\textbf {\bibinfo {volume} {88}},\ \bibinfo {pages} {064408}
  (\bibinfo {year} {2013})}\BibitemShut {NoStop}%
\bibitem [{\citenamefont {{Agrawal}}\ \emph {et~al.}()\citenamefont
  {{Agrawal}}, \citenamefont {{Vasyuchka}}, \citenamefont {{Serga}},
  \citenamefont {{Kirihara}}, \citenamefont {{Pirro}}, \citenamefont
  {{Langner}}, \citenamefont {{Jungfleisch}}, \citenamefont {{Chumak}},
  \citenamefont {{Papaioannou}},\ and\ \citenamefont {{Hillebrands}}}]{ref31}%
  \BibitemOpen
  \bibfield  {author} {\bibinfo {author} {\bibfnamefont {M.}~\bibnamefont
  {{Agrawal}}}, \bibinfo {author} {\bibfnamefont {V.~I.}\ \bibnamefont
  {{Vasyuchka}}}, \bibinfo {author} {\bibfnamefont {A.~A.}\ \bibnamefont
  {{Serga}}}, \bibinfo {author} {\bibfnamefont {A.}~\bibnamefont {{Kirihara}}},
  \bibinfo {author} {\bibfnamefont {P.}~\bibnamefont {{Pirro}}}, \bibinfo
  {author} {\bibfnamefont {T.}~\bibnamefont {{Langner}}}, \bibinfo {author}
  {\bibfnamefont {M.~B.}\ \bibnamefont {{Jungfleisch}}}, \bibinfo {author}
  {\bibfnamefont {A.~V.}\ \bibnamefont {{Chumak}}}, \bibinfo {author}
  {\bibfnamefont {E.~T.}\ \bibnamefont {{Papaioannou}}}, \ and\ \bibinfo
  {author} {\bibfnamefont {B.}~\bibnamefont {{Hillebrands}}},\ }\href@noop {}
  {\ }\Eprint {http://arxiv.org/abs/1309.2164} {arXiv:1309.2164} \BibitemShut
  {NoStop}%
\bibitem [{\citenamefont {Uchida}\ \emph
  {et~al.}(2010{\natexlab{b}})\citenamefont {Uchida}, \citenamefont {Adachi},
  \citenamefont {Ota}, \citenamefont {Nakayama}, \citenamefont {Maekawa},\ and\
  \citenamefont {Saitoh}}]{ref28}%
  \BibitemOpen
  \bibfield  {author} {\bibinfo {author} {\bibfnamefont {K.}~\bibnamefont
  {Uchida}}, \bibinfo {author} {\bibfnamefont {H.}~\bibnamefont {Adachi}},
  \bibinfo {author} {\bibfnamefont {T.}~\bibnamefont {Ota}}, \bibinfo {author}
  {\bibfnamefont {H.}~\bibnamefont {Nakayama}}, \bibinfo {author}
  {\bibfnamefont {S.}~\bibnamefont {Maekawa}}, \ and\ \bibinfo {author}
  {\bibfnamefont {E.}~\bibnamefont {Saitoh}},\ }\href {\doibase
  10.1063/1.3507386} {\bibfield  {journal} {\bibinfo  {journal} {Appl. Phy.
  Lett.}\ }\textbf {\bibinfo {volume} {97}},\ \bibinfo {pages} {172505}
  (\bibinfo {year} {2010}{\natexlab{b}})}\BibitemShut {NoStop}%
\bibitem [{\citenamefont {Weiler}\ \emph {et~al.}(2012)\citenamefont {Weiler},
  \citenamefont {Althammer}, \citenamefont {Czeschka}, \citenamefont {Huebl},
  \citenamefont {Wagner}, \citenamefont {Opel}, \citenamefont {Imort},
  \citenamefont {Reiss}, \citenamefont {Thomas}, \citenamefont {Gross},\ and\
  \citenamefont {Goennenwein}}]{ref29}%
  \BibitemOpen
  \bibfield  {author} {\bibinfo {author} {\bibfnamefont {M.}~\bibnamefont
  {Weiler}}, \bibinfo {author} {\bibfnamefont {M.}~\bibnamefont {Althammer}},
  \bibinfo {author} {\bibfnamefont {F.~D.}\ \bibnamefont {Czeschka}}, \bibinfo
  {author} {\bibfnamefont {H.}~\bibnamefont {Huebl}}, \bibinfo {author}
  {\bibfnamefont {M.~S.}\ \bibnamefont {Wagner}}, \bibinfo {author}
  {\bibfnamefont {M.}~\bibnamefont {Opel}}, \bibinfo {author} {\bibfnamefont
  {I.-M.}\ \bibnamefont {Imort}}, \bibinfo {author} {\bibfnamefont
  {G.}~\bibnamefont {Reiss}}, \bibinfo {author} {\bibfnamefont
  {A.}~\bibnamefont {Thomas}}, \bibinfo {author} {\bibfnamefont
  {R.}~\bibnamefont {Gross}}, \ and\ \bibinfo {author} {\bibfnamefont
  {S.~T.~B.}\ \bibnamefont {Goennenwein}},\ }\href {\doibase
  10.1103/PhysRevLett.108.106602} {\bibfield  {journal} {\bibinfo  {journal}
  {Phys. Rev. Lett.}\ }\textbf {\bibinfo {volume} {108}},\ \bibinfo {pages}
  {106602} (\bibinfo {year} {2012})}\BibitemShut {NoStop}%
\bibitem [{\citenamefont {Qu}\ \emph {et~al.}(2013)\citenamefont {Qu},
  \citenamefont {Huang}, \citenamefont {Hu}, \citenamefont {Wu},\ and\
  \citenamefont {Chien}}]{ref26}%
  \BibitemOpen
  \bibfield  {author} {\bibinfo {author} {\bibfnamefont {D.}~\bibnamefont
  {Qu}}, \bibinfo {author} {\bibfnamefont {S.~Y.}\ \bibnamefont {Huang}},
  \bibinfo {author} {\bibfnamefont {J.}~\bibnamefont {Hu}}, \bibinfo {author}
  {\bibfnamefont {R.}~\bibnamefont {Wu}}, \ and\ \bibinfo {author}
  {\bibfnamefont {C.~L.}\ \bibnamefont {Chien}},\ }\href {\doibase
  10.1103/PhysRevLett.110.067206} {\bibfield  {journal} {\bibinfo  {journal}
  {Phys. Rev. Lett.}\ }\textbf {\bibinfo {volume} {110}},\ \bibinfo {eid}
  {067206} (\bibinfo {year} {2013})}\BibitemShut {NoStop}%
\bibitem [{\citenamefont {{Kikkawa}}\ \emph {et~al.}(2013)\citenamefont
  {{Kikkawa}}, \citenamefont {{Uchida}}, \citenamefont {{Shiomi}},
  \citenamefont {{Qiu}}, \citenamefont {{Hou}}, \citenamefont {{Tian}},
  \citenamefont {{Nakayama}}, \citenamefont {{Jin}},\ and\ \citenamefont
  {{Saitoh}}}]{ref27}%
  \BibitemOpen
  \bibfield  {author} {\bibinfo {author} {\bibfnamefont {T.}~\bibnamefont
  {{Kikkawa}}}, \bibinfo {author} {\bibfnamefont {K.}~\bibnamefont {{Uchida}}},
  \bibinfo {author} {\bibfnamefont {Y.}~\bibnamefont {{Shiomi}}}, \bibinfo
  {author} {\bibfnamefont {Z.}~\bibnamefont {{Qiu}}}, \bibinfo {author}
  {\bibfnamefont {D.}~\bibnamefont {{Hou}}}, \bibinfo {author} {\bibfnamefont
  {D.}~\bibnamefont {{Tian}}}, \bibinfo {author} {\bibfnamefont
  {H.}~\bibnamefont {{Nakayama}}}, \bibinfo {author} {\bibfnamefont {X.-F.}\
  \bibnamefont {{Jin}}}, \ and\ \bibinfo {author} {\bibfnamefont
  {E.}~\bibnamefont {{Saitoh}}},\ }\href {\doibase
  10.1103/PhysRevLett.110.067207} {\bibfield  {journal} {\bibinfo  {journal}
  {Phys. Rev. Lett.}\ }\textbf {\bibinfo {volume} {110}},\ \bibinfo {eid}
  {067207} (\bibinfo {year} {2013})}\BibitemShut {NoStop}%
\end{thebibliography}%

\end{document}